\documentclass[letterpaper, 10 pt, conference]{ieeeconf}  

\IEEEoverridecommandlockouts                              

\usepackage{graphicx}

\usepackage{marvosym}  

\usepackage{booktabs} 
\usepackage{multirow} 
\usepackage{amsmath}  
\title{
\large \bf
CTAN: Cycle-Temporal Attention Network for Embodied Audio-Visual Navigation$\dagger$
}

\author{
Teng Liu$^{1,2,3}$, and Yinfeng Yu$^{1,2,3}$$^{,\mbox{\Letter}}$%
\thanks{\small $\dagger$This research was financially supported by the National Natural Science Foundation of China (Grant No. 62463029).}
\thanks{\small $^1$School of Computer Science and Technology, Xinjiang University, Urumqi 830017, China.}
\thanks{\small $^2$Joint Research Laboratory for Embodied Intelligence, Xinjiang University.}
\thanks{\small $^3$Joint International Research Laboratory of Silk Road Multilingual Cognitive Computing, Xinjiang University.}
\thanks{\small $^{\mbox{\Letter}}$Yinfeng Yu is the corresponding author (Email: yuyinfeng@xju.edu.cn).}%
}

\begin{document}

\maketitle
\thispagestyle{empty}
\pagestyle{empty}

\begin{abstract}
Audio-visual embodied navigation equips robots with the capability to infer the locations of sound sources by integrating visual inputs and acoustic information (e.g., depth observations and binaural audio cues). The core challenge lies in establishing effective semantic interactions across heterogeneous modalities (which exhibit distinct feature distributions). Existing feature fusion strategies, however, often rely on simple multimodal aggregation and therefore fail to capture the underlying geometric and semantic relationships, leading to information degradation in complex environments.
To overcome these limitations, this work presents the Cycle-Temporal Attention Network (CTAN), a framework designed for active semantic-enhanced fusion (rather than straightforward multimodal combination). Specifically, the proposed Audio-Visual Reconstruction Cross-Attention (AVRCA) module employs a bidirectional cycle-consistency constraint (between visual and acoustic representations) to reinforce the spatial semantic attributes of both modalities, thereby facilitating more robust cross-modal interaction. Additionally, we design a Temporal Cross-Modal Memory (TCMM) mechanism to dynamically integrate real-time enhanced multimodal features with historical context, reducing performance drops caused by auditory dead zones. Experimental results obtained on the Replica and Matterport3D benchmarks indicate that the proposed approach achieves superior performance over previous audio-visual navigation methods in terms of success rate (SR), success weighted by path length (SPL), and scene navigation accuracy (SNA).

\end{abstract}

\section{INTRODUCTION}

Embodied intelligence~\cite{yu2025dynamic,yang2026beyond,yu2025dope} has been widely studied across various research fields, among which audio-visual navigation~\cite{chen2020soundspaces} has emerged as an important direction. In complex 3D environments, audio-visual navigation~\cite{chen2020soundspaces} enables autonomous agents to perform target localization and path planning with auditory cues, an essential capability for embodied agents.

Humans naturally integrate visual and auditory information during navigation: sound provides coarse spatial guidance, while vision enables fine-grained localization~\cite{yu2025dgfnet,fu2025fsdenet,mattursun2024bss}. However, existing agents still struggle to achieve such natural and effective cross-modal collaboration. Most methods rely on simple feature combinations, which cannot effectively capture deep semantic connections between vision and audio. As a result, complementary information across heterogeneous modalities is underutilized, degrading navigation performance in complex scenarios. Furthermore, such direct fusion strategies fail to maintain durable semantic interactions, especially when individual modalities are disturbed or incomplete, making stable perceptual consistency difficult to sustain~\cite{wang2025modality,cao2024vnet,zhang2024nonlinear}. These limitations indicate that naive cross-modal combination is insufficient for reliable and robust navigation.

To overcome these challenges, we develop the Cycle-Temporal Attention Network (CTAN). The framework introduces an Audio-Visual Reconstruction Cross-Attention (AVRCA) module to enhance the mutual semantics of visual and audio features via bidirectional cycle consistency. Based on this, a Temporal Cross-Modal Memory (TCMM) mechanism is incorporated to integrate historical context and maintain continuous perceptual representation. The proposed method effectively constructs stable semantic associations across modalities, improving the generalization and robustness of audio-visual navigation in complex environments.

\begin{itemize}\item We propose the Cycle-Temporal Attention Network (CTAN), which integrates the Audio-Visual Reconstruction Cross-Attention (AVRCA) and Temporal Cross-Modal Memory (TCMM) for deep cross-modal fusion. AVRCA achieves bidirectional semantic enhancement through a symmetric cycle-consistency objective, enabling visual and acoustic features to mutually reinforce each other at the semantic level. TCMM stabilizes this enhancement process by adaptively fusing the semantically enhanced perceptions with historical context, helping to bridge auditory dead zones and ensure persistent spatial awareness.
\item Our approach also highlights the adaptive dynamics between instantaneous semantically enhanced perceptions and long-term historical context during navigation. These temporal interactions help construct a persistent spatial representation, effectively mitigating the impact of auditory dead zones and ensuring stable, coherent policy execution in complex indoor environments.
\item The proposed framework is evaluated on the challenging 3D benchmarks Replica and Matterport3D. Results indicate that CTAN consistently surpasses existing cross-modal fusion methods in navigation efficiency and success rate, demonstrating its effectiveness.

\end{itemize}

\begin{figure*}[htbp]
    \centering
   
    \includegraphics[width=0.95\textwidth]{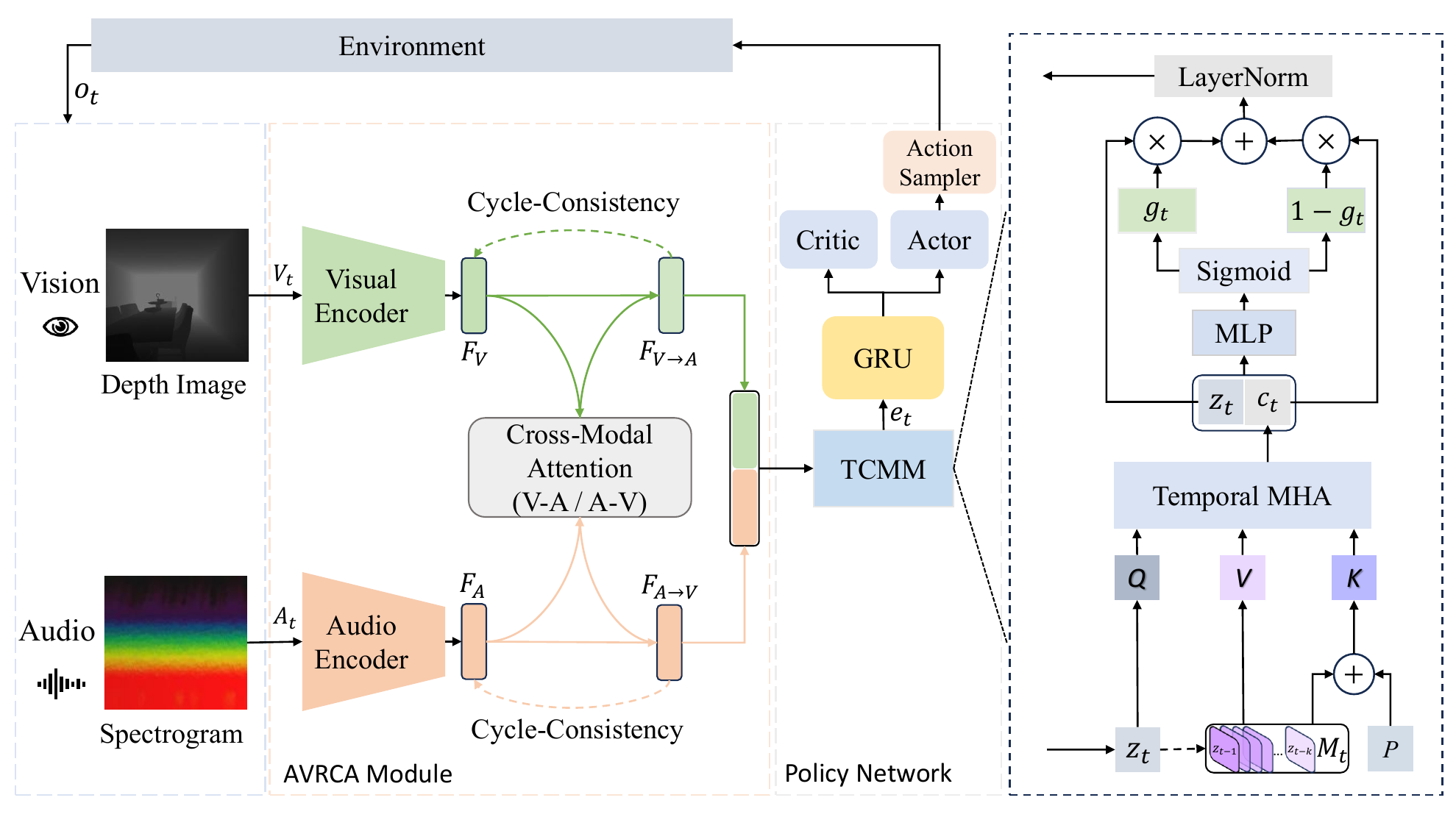} 
    \caption{Given visual depth images and audio spectrograms as input, CTAN first extracts modality-specific features, then performs bidirectional semantic enhancement via the Audio-Visual Reconstruction Cross-Attention (AVRCA) module with cycle-consistency regularization. The Temporal Cross-Modal Memory (TCMM) module subsequently integrates historical context to maintain stable spatial perception, which is finally fed into an actor-critic framework for sound-guided navigation policy learning.
}
    \label{fig:fra}
\end{figure*}

\section{RELATED WORK}

Audio-Visual Navigation (AVN) has been significantly promoted by the SoundSpaces framework~\cite{chen2020soundspaces}, which leverages real-scanned environments, including Replica~\cite{straub2019replica} and Matterport3D~\cite{chang2017matterport3d}, to enable high-fidelity acoustic simulation for embodied agents. Early studies within this benchmark mainly explored the connection between multimodal perception and navigation policies. LLA~\cite{gan2020look} proposed a hierarchical design for more stable policy optimization, whereas AV-WAN~\cite{chen2020learning} and CAHM~\cite{younes2023catch} incorporated structured spatial representations (e.g., audio-visual maps and local geometric memory) to capture long-term environmental dependencies. These approaches demonstrated that auditory information provides valuable guidance for improving navigation performance in unseen 3D environments.

As research advances, successive works explore richer contextual modeling for complex navigation environments. SAVi~\cite{chen2021semantic} integrates semantic label embedding (establishing explicit correspondence between sound events and visual objects) for cross-modal association. AVLEN~\cite{paul2022avlen} and CAVEN~\cite{liu2024caven} adopt external oracle guidance (introducing prior multimodal knowledge for audio-visual feature alignment). For better generalization in challenging scenarios, SAAVN~\cite{YinfengICLR2022saavn} uses adversarial learning (enhancing disturbance robustness), while ORAN~\cite{chen2023omnidirectional} employs omnidirectional perception and distillation (overcoming single-view limitations).
However, most existing methods rely on passive and shallow cross-modal fusion (neglecting the inherent spatial and physical coupling between audio and visual signals)~\cite{li2025audio,zhang2025advancing,yu2023measuring}. In contrast, the proposed CTAN applies bidirectional cycle consistency (realizing active cross-modal semantic enhancement) to capture invariant physical priors, enabling more robust agent navigation in unknown environments~\cite{shah2019cycle}.

Temporal modeling plays an important role in audio-visual navigation when acoustic observations become unreliable. Existing approaches often rely on recurrent networks or transformers for historical information aggregation, yet they may fail to recover useful context in auditory dead zones (such as areas affected by sound occlusion or attenuation). To address this issue, TCMM adopts a dynamic context buffer (with adaptive selection of historical features) to preserve temporal continuity and alleviate navigation errors in complex indoor scenes.

\section{METHODOLOGY}
\subsection{Problem Formalization}
The task is formulated within a reinforcement learning framework, where the robot acquires a policy for efficient sound source localization in unknown 3D environments. The CTAN (Cycle-Temporal Attention Network) framework consists of four key modules, as shown in Fig.~\ref{fig:fra}. Given the egocentric depth observations and acoustic spectrograms as inputs, our model 1) extracts multimodal features through specialized visual and audio encoders; 2) performs bidirectional semantic enhancement through the Audio-Visual Reconstruction Cross-Attention (AVRCA) module, which leverages cycle consistency to generate physically consistent embeddings; 3) integrates these instantaneous perceptions with historical context through the Temporal Cross-Modal Memory (TCMM) module to maintain spatial awareness; and finally 4) employs an actor-critic architecture, supported by a GRU-based state representation, to perform action prediction and policy optimization. The robot agent repeats this iterative process of perception and decision-making until the target sound source is successfully reached. We will introduce each component consecutively in the following sections.

\subsection{Cross-Modal Fusion}

At each decision step $t$, the agent observes an audio-visual input $O_t=(V_t,A_t)$. The depth image $V_t$ and binaural spectrogram $A_t$ are encoded separately using two CNN encoders to obtain visual and acoustic representations. Each encoder adopts an identical architecture consisting of three convolutional layers (Conv8$\times$8, Conv4$\times$4, Conv3$\times$3), followed by a fully connected layer with 512 output units and ReLU activation after every layer. The extracted visual and acoustic features are denoted by $F_V \in \mathbf{R}^{d}$ and $F_A \in \mathbf{R}^{d}$, respectively, where $d=512$ in all experiments.

To overcome the limitations of passive feature aggregation, we introduce the Audio-Visual Reconstruction Cross-Attention (AVRCA) module. The core of AVRCA is a Cross-Modal Attention (CMA) mechanism that facilitates bidirectional semantic enhancement by treating each modality as a query to probe the other. As shown in the center of Fig.~\ref{fig:fra}, instead of direct concatenation, we first perform bidirectional cross-modal translation. By utilizing $F_V$ as the query to attend to the audio key-value pairs, we obtain the audio-augmented visual feature $F_{V \to A}$, which adaptively aggregates relevant audio context for each visual patch:
\begin{equation}
    F_{V \to A} = \mathrm{CrossAttn}(Q=F_V, K=F_A, V=F_A),
\end{equation}
Symmetrically, the visual-augmented audio feature $F_{A \to V}$ is computed as:
\begin{equation}
    F_{A \to V} = \mathrm{CrossAttn}(Q=F_A, K=F_V, V=F_V).
\end{equation}

Relying solely on cross-attention may lead to trivial solutions or attention collapse. To mathematically guarantee that the translated features retain the essential modality-specific structures, we introduce a self-supervised cycle-consistency constraint. Specifically, we use the translated features as queries to reconstruct the original modality representations:
\begin{equation}
    \tilde{F}_V = \mathrm{CrossAttn}(Q=F_{V \to A}, K=F_V, V=F_V),
\end{equation}
\begin{equation}
    \tilde{F}_A = \mathrm{CrossAttn}(Q=F_{A \to V}, K=F_A, V=F_A),
\end{equation}
The cycle-consistency loss $\mathcal{L}_{cycle}$ is defined as the Mean Squared Error (MSE) between the original feature representations and their cyclic reconstructions:
\begin{equation}
    \mathcal{L}_{cycle} = \frac{1}{N} \left( \Vert F_V - \tilde{F}_V \Vert^2_2 + \Vert F_A - \tilde{F}_A \Vert^2_2 \right),
\end{equation}
This loss acts as a strict structural regularizer during training. It forces the network to bind the sounding object with its corresponding acoustic signature to prevent information loss during the translation cycle, ensuring strict semantic consistency in the latent space.

Finally, the enhanced modality representations are fused through concatenation and mapped by an MLP, producing the spatial representation for the current time step:
\begin{equation}
    z_t = \mathrm{MLP}([F_{V \to A}, F_{A \to V}]),
\end{equation}
This context-rich vector $z_t$ acts as the input for the subsequent temporal sequence modeling in the TCMM module.

\subsection{Adaptive Temporal Reasoning}

The fused embedding $z_t \in \mathbf{R}^{2d}$ is further processed by the Temporal Cross-Modal Memory (TCMM) module. As shown on the right side of Fig.~\ref{fig:fra}, TCMM maintains a sliding historical memory $M_t = \{z_{t-k}, \dots, z_{t-1}\} \in \mathbf{R}^{k \times 2d}$, where $k$ denotes the temporal window size ($k=10$ in our implementation).

To extract relevant cues from past observations, we employ a Temporal Multi-Head Attention (MHA) mechanism. Specifically, we use the current perception $z_t$ as the query ($Q$), while the keys ($K$) and values ($V$) are derived from the memory bank $M_t$ augmented with sinusoidal positional encodings $P \in \mathbf{R}^{k \times 2d}$ to preserve temporal order. The contextual cue $c_t \in \mathbf{R}^{2d}$ is computed as:
\begin{equation}
c_t = \mathrm{MHA}(Q=z_t, K=M_t+P, V=M_t).
\end{equation}

To determine the relative importance of immediate perceptions versus historical context in real-time, we design an Adaptive Gating Mechanism. We first concatenate the current embedding $Z_t$ and the contextual cue $c_t$ to compute a gating vector $g_t \in \mathbf{R}^{2d}$ via a Multi-Layer Perceptron (MLP) and a Sigmoid activation function:
\begin{equation}
g_t = \sigma(\mathrm{MLP}([z_t; c_t])),
\end{equation}
Here, $\sigma$ represents the Sigmoid activation, while $[\cdot ; \cdot]$ denotes feature concatenation. This gate adaptively balances the features to form the temporal-aware state $\hat{e}_t$:
\begin{equation}
\hat{e}_t = g_t \odot z_t + (1 - g_t) \odot c_t,
\end{equation}
where $\odot$ denotes the element-wise product. The final state representation $s_t$ is obtained after applying Layer Normalization to ensure training stability:
\begin{equation}
e_t = \mathrm{LayerNorm}(\hat{e}_t).
\end{equation}

In the final stage, $e_t$ is fed into a GRU with a 512-dim hidden layer, followed by an Actor-Critic network for policy learning and value estimation.

\begin{table*}[htbp]
\centering
\small 
\setlength{\tabcolsep}{5.5pt} 

\caption{Comparison of navigation performance under the Depth setting. SPL, SR, and SNA are reported as percentages.}
\label{tab:main}
\renewcommand{\arraystretch}{1.3} 
\begin{tabular}{l | c c c | c c c | c c c | c c c}
\hline
 & \multicolumn{6}{c|}{{Replica}} & \multicolumn{6}{c}{{Matterport3D}} \\
\cline{2-13}
{Model} & \multicolumn{3}{c|}{{Multiple heard}} & \multicolumn{3}{c|}{{Multiple unheard}} & \multicolumn{3}{c|}{{Multiple heard}} & \multicolumn{3}{c}{{Multiple unheard}} \\
\cline{2-13}
 & {SPL$\uparrow$} & {SR$\uparrow$} & {SNA$\uparrow$} & {SPL$\uparrow$} & {SR$\uparrow$} & {SNA$\uparrow$} & {SPL$\uparrow$} & {SR$\uparrow$} & {SNA$\uparrow$} & {SPL$\uparrow$} & {SR$\uparrow$} & {SNA$\uparrow$} \\
\hline
Random Agent~\cite{chen2020learning} & 4.9 & 18.5 & 1.8 & 4.9 & 18.5 & 1.8 & 2.1 & 9.1 & 0.8 & 2.1 & 9.1 & 0.8 \\
Dir. Follower~\cite{chen2020learning} & 54.7 & 72.0 & 41.1 & 11.1 & 17.2 & 8.4 & 32.3 & 41.2 & 23.8 & 13.9 & 18.0 & 10.7 \\
Frontier Waypoints~\cite{chen2020learning} & 44.0 & 63.9 & 35.2 & 6.5 & 14.8 & 5.1 & 30.6 & 42.8 & 22.2 & 10.9 & 16.4 & 8.1 \\
Supervised Waypoints~\cite{chen2020learning} & 59.1 & 88.1 & 48.5 & 14.1 & 43.1 & 10.1 & 21.0 & 36.2 & 16.2 & 4.1 & 8.8 & 2.9 \\
Gan et al.~\cite{gan2020look} & 57.6 & 83.1 & 47.9 & 7.5 & 15.7 & 5.7 & 22.8 & 37.9 & 17.1 & 5.0 & 10.2 & 3.6 \\
AGSA~\cite{li2025audio} & 75.5 & 93.2 & 52.0 & 36.6 & 48.3 & 22.4 & 54.1 & 70.0 & 30.0 & 36.5 & 26.2 & 13.1 \\
SoundSpaces~\cite{chen2020soundspaces} & 76.1 & 93.0 & 44.7 & 35.8 & 47.0 & 20.9 & 52.3 & 68.8 & \textbf{29.6} & 21.9 & 33.5 & 10.4 \\
\textbf{CTAN(Ours)} & \textbf{78.6} & \textbf{94.1} & \textbf{49.7} & \textbf{40.7} & \textbf{53.3} & \textbf{23.3} & \textbf{54.2} & \textbf{71.4} & 29.0 & \textbf{27.4} & \textbf{37.7} & \textbf{14.2} \\
\hline
\end{tabular}
\end{table*}

\section{EXPERIMENTS}
\subsection{Experimental Setup}
\textbf{Datasets.} Experiments are conducted on the Replica~\cite{straub2019replica} and Matterport3D~\cite{chang2017matterport3d} benchmarks to assess the proposed CTAN. Replica contains 18 high-fidelity reconstructed 3D scenes, providing accurate depth information, camera poses, and detailed mesh models. Matterport3D includes 90 real-world indoor environments collected at the building scale, with an average area of approximately 517 square meters. To ensure a consistent evaluation, we adopt the experimental settings of SoundSpaces~\cite{chen2020soundspaces}, using its widely adopted 85-scene subset together with the official train/validation/test partition~\cite{zhang2025iterative}. 

\textbf{Experimental Environment.}  
We instantiate our experimental setup within the Habitat simulator, utilizing the SoundSpaces framework for high-fidelity audio rendering. Experiments are conducted in the Replica~\cite{straub2019replica} and Matterport3D~\cite{chang2017matterport3d} environments. To achieve realistic acoustic simulation, the environment incorporates room impulse responses (RIRs) and material-specific properties generated via bidirectional path-tracing. This configuration yields a spatially and acoustically accurate 3D space, enabling the agent to learn robust sound-guided navigation policies for unseen environments.
\begin{table}[ht]
\centering
\caption{Ablation study evaluating the effectiveness of individual components on the Replica dataset.}
\label{tab:rep}
\footnotesize
\setlength{\tabcolsep}{3.5pt} 
\renewcommand{\arraystretch}{1.3} 
\begin{tabular}{l|ccc|ccc}
\hline
 & \multicolumn{6}{c}{Replica} \\
\cline{2-7}
Model & \multicolumn{3}{c|}{Multiple heard} & \multicolumn{3}{c}{Multiple unheard} \\
\cline{2-7}
 & SPL($\uparrow$) & SR($\uparrow$) & SNA($\uparrow$) & SPL($\uparrow$) & SR($\uparrow$) & SNA($\uparrow$) \\
\hline
w/o AVRCA & 75.0 & 92.2 & 48.5& 36.3 & 46.8 & 21.6 \\
w/o TCMM  & 75.7 & 92.5 & 48.2 & 36.7 & 47.5 & 21.5 \\
w/o $L_{\mathrm{cyc}}$ & 75.6 & 93.0 & 49.3 & 36.9 & 47.7 & 21.6 \\ 
\textbf{CTAN(Ours)} & \textbf{78.6} & \textbf{94.1} & \textbf{49.7} & \textbf{40.7} & \textbf{53.3} & \textbf{23.3} \\
\hline
\multicolumn{7}{l}\textit{Note:} ``w/o'' indicates that the corresponding component is removed. \\
\end{tabular}
\end{table}

\begin{table}[ht]
\centering
\caption{Ablation study evaluating the effectiveness of individual components on the Matterport3D dataset.}
\label{tab:mp}
\footnotesize
\setlength{\tabcolsep}{3.5pt} 
\renewcommand{\arraystretch}{1.3} 
\begin{tabular}{l|ccc|ccc}
\hline
 & \multicolumn{6}{c}{Matterport3D} \\
\cline{2-7}
Model & \multicolumn{3}{c|}{Multiple heard} & \multicolumn{3}{c}{Multiple unheard} \\
\cline{2-7}
 & SPL($\uparrow$) & SR($\uparrow$) & SNA($\uparrow$) & SPL($\uparrow$) & SR($\uparrow$) & SNA($\uparrow$) \\
\hline
w/o AVRCA & 50.9 & 69.4 & 27.0 & 23.9 & 35.3 & 12.5 \\
w/o TCMM & 53.7 & 70.3 & 28.0 & 21.9 & 33.5 & 10.4 \\
w/o $L_{\mathrm{cyc}}$ & 53.9 & 67.3 & \textbf{30.2}& 26.1 & 36.8 & 13.7 \\ 
\textbf{CTAN(Ours)} & \textbf{54.2} & \textbf{71.4} & {29.0} & \textbf{27.4} & \textbf{37.7} & \textbf{14.2} \\
\hline
\multicolumn{7}{l}\textit{Note:} ``w/o'' indicates that the corresponding component is removed. \\
\end{tabular}
\end{table}
\textbf{Evaluation Metrics.} Performance is assessed using three commonly adopted metrics~\cite{chen2020soundspaces}: (1) Success Rate (SR), representing the percentage of episodes in which the agent reaches the target within the allowed number of steps; (2) Success weighted by Path Length (SPL), reflecting navigation efficiency by relating the executed path to the shortest achievable route; and (3) Success-Navigation Accuracy (SNA), measuring both successful goal attainment and the accuracy of the agent's final orientation toward the sound source.

\subsection{Performance Comparison}
We follow the SoundSpaces evaluation protocol~\cite{chen2020soundspaces} and consider two sound generalization settings: (1) \textit{Multiple heard sounds}, where the training and testing phases involve the same sound categories, and (2) \textit{Multiple unheard sounds}, where the test sounds are not observed during training. The evaluation environments remain unseen in both cases (eliminating the possibility of relying on memorized spatial layouts) and therefore better reflect cross-modal generalization ability.

Table~I reports the quantitative comparison. CTAN obtains superior results on most evaluation metrics. Compared with SoundSpaces~\cite{chen2020soundspaces}, SPL increases by 2.5\% and 1.9\% for heard sounds, and by 4.9\% and 5.5\% for unheard sounds on Replica and Matterport3D, respectively (showing more efficient path planning). SR also improves by 1.1\% and 2.6\% in the heard setting and by 6.3\% and 4.2\% in the unheard setting (where unseen acoustic conditions introduce additional difficulties), highlighting the robustness of CTAN in complex environments.
\begin{figure*}[htbp]
    \centering
 
    \includegraphics[width=0.95\textwidth]{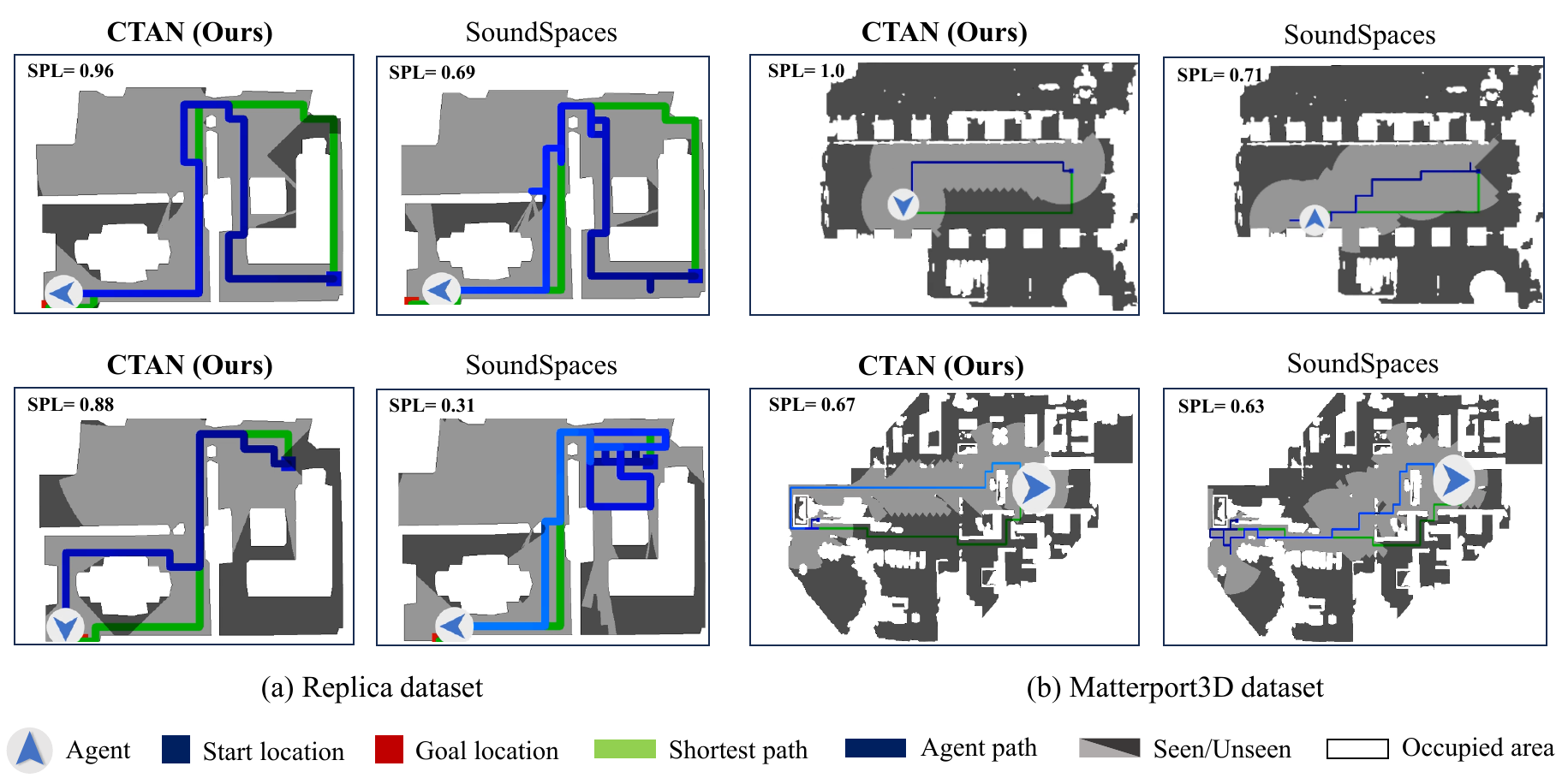} 
    \caption{Navigation trajectories at the end of representative episodes in (a) Replica and (b) Matterport3D. Higher SPL values and shorter blue trajectories correspond to better navigation performance.}
    \label{fig:tra}
\end{figure*}
\subsection{Ablation Study}
Ablation experiments are carried out on the Replica and Matterport3D datasets to examine the role of each major component in CTAN. Three modules—AVRCA, TCMM, and the cycle-consistency loss ($\mathcal{L}_{cyc}$)—are removed individually (one at a time) to evaluate their influence on navigation performance. As reported in Tables~\ref{tab:rep} and~\ref{tab:mp}, removing any component generally degrades performance under both the Heard and Unheard settings (with different levels of impact), indicating that each module contributes to the effectiveness of CTAN.

Removing AVRCA causes a noticeable decline in SR and SPL, especially under the Unheard setting on Replica (where SR drops by 6.5\%). This observation highlights the importance of cross-modal semantic enhancement (particularly when encountering novel sound sources). Conversely, eliminating TCMM mainly affects the Unheard setting on Matterport3D, reducing SPL from 27.4\% to 21.9\% (a larger decrease than under the Heard setting). The result suggests that temporal memory is particularly valuable in large-scale environments (preserving historical context over long navigation trajectories).

Omitting the cycle-consistency loss ($\mathcal{L}_{cyc}$) also reduces navigation efficiency. Although the performance decline is smaller than that caused by removing the architectural modules, SPL decreases consistently on both datasets (regardless of the evaluation setting). This indicates that the cyclic constraint serves as an effective regularizer (reinforcing stable audio-visual associations), rather than allowing the model to rely mainly on spatial memorization.
\subsection{Qualitative Analysis}
\textbf{Navigation trajectories.}
Fig.~\ref{fig:tra} illustrates representative top-down navigation trajectories of CTAN and SoundSpaces on the Replica and Matterport3D datasets. Across the sampled episodes, CTAN generally produces routes that remain closer to the theoretical shortest paths (as reflected by higher SPL values), whereas SoundSpaces is more prone to inefficient exploration (e.g., local wandering or repeated backtracking), leading to larger path deviations. These trajectory comparisons indicate that combining cross-modal semantic enhancement with temporal memory (rather than relying on simple feature concatenation) improves navigation efficiency in complex environments.

\begin{figure*}[htbp]
    \centering

    \includegraphics[width=0.95\textwidth]{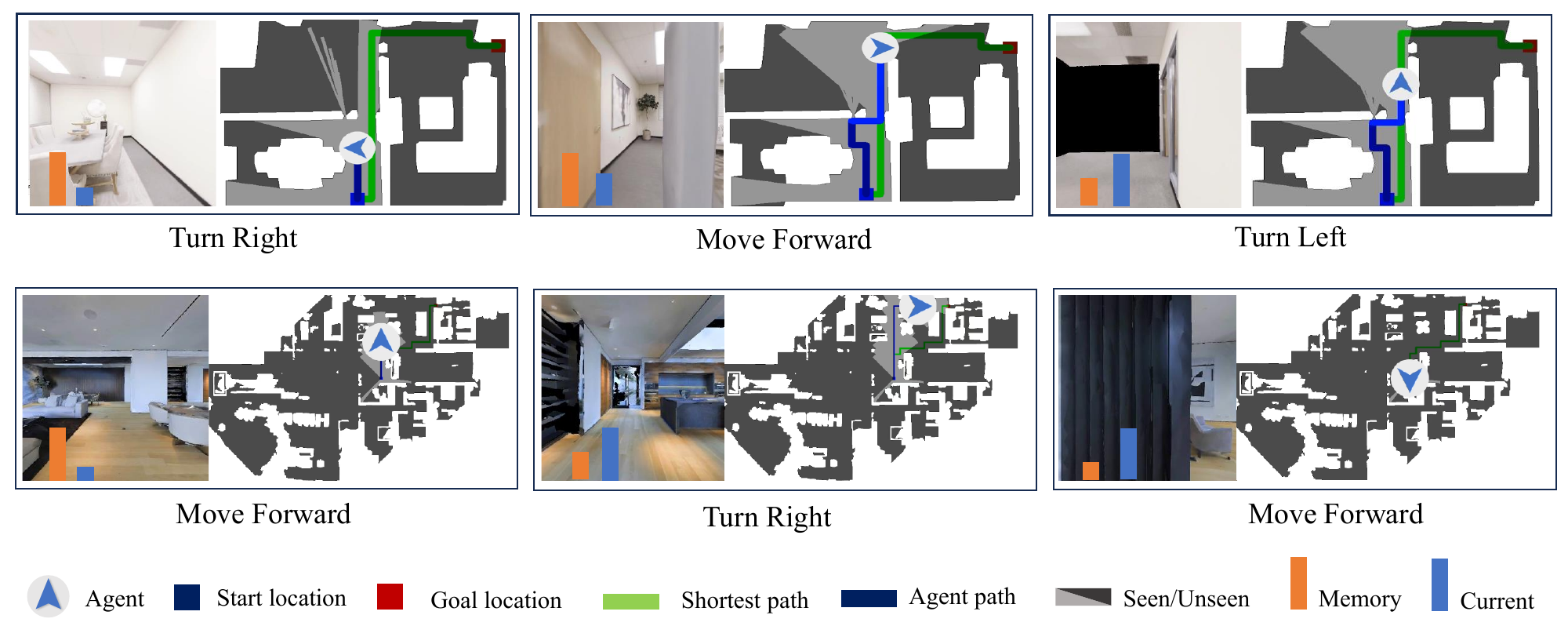} 
    \caption{Dynamic contribution of memory and current perception for two episodes. Columns display feature importance at three sampled time steps. The orange and blue bars represent the relative weights of historical memory and immediate perception, respectively.}
    \label{fig:dy}
\end{figure*}

\textbf{Dynamic Gating Analysis.}
During audio-visual navigation, the reliance on current observations and historical context varies as the agent moves through complex environments. Fig.~\ref{fig:dy} presents the normalized memory and perception weights obtained from the TCMM gating values ($g_t$ and $1-g_t$), visualized as bar charts. The weight distribution at different action steps reflects the changing roles of recent observations and accumulated experience (instead of using a fixed memory strategy). 
\section{CONCLUSIONS}


We develop CTAN (Cycle-Temporal Attention Network) to improve audio-visual embodied navigation in complex 3D environments, where effective cross-modal interaction and temporal perception remain challenging (especially when acoustic observations are incomplete). The framework introduces two complementary designs: the Audio-Visual Reconstruction Cross-Attention (AVRCA) module, which enhances the semantic correspondence between visual and auditory features, and the Temporal Cross-Modal Memory (TCMM) mechanism, which maintains useful historical information during navigation.

Evaluation on Replica and Matterport3D shows that CTAN achieves consistent improvements in navigation success, path efficiency, and scene adaptability, particularly for unseen sounds and large-scale indoor environments (where reliable perception is more difficult). Beyond the current setting, extending the framework to dynamic scenarios and incorporating richer sensory information (such as tactile or language cues) provide promising directions for future embodied intelligence systems.


\bibliographystyle{IEEEtran} 
\bibliography{ref}           

\end{document}